# Microwave Background Temperature at a Redshift of 6.34 from $H_2O$ Absorption

Dominik A. Riechers[1], Axel Weiss[2], Fabian Walter[3,4], Christopher L. Carilli[4], Pierre Cox[5], Roberto Decarli[6], and Roberto Neri[7]

[1] I. Physikalisches Institut, Universität zu Köln, Zülpicher Str. 77, D-50937 Köln, Germany
[2] Max-Planck-Institut für Radioastronomie, Auf dem Hügel 69, D-53121 Bonn, Germany
[3] Max-Planck-Institut für Astronomie, Königstuhl 17, D-69117 Heidelberg, Germany
[4] National Radio Astronomy Observatory, Pete V. Domenici Array Science Center, P.O. Box O, Socorro, NM 87801, USA
[5] Sorbonne Université, UPMC Université Paris 6 and CNRS, UMR 7095, Institut d'Astrophysique de Paris, 98bis boulevard Arago, F-75014 Paris, France
[6] INAF - Osservatorio di Astrofisica e Scienza dello Spazio, via Gobetti 93/3, I-40129, Bologna, Italy
[7] Institut de RadioAstronomie Millimétrique, 300 Rue de la Piscine, Domaine Universitaire, F-38406 Saint Martin d'Héres, France

**Distortions of the observed cosmic microwave background imprinted by the Sunyaev-Zel'dovich effect toward massive galaxy clusters due to inverse Compton scattering of microwave photons by high-energy electrons provide a direct measurement of the microwave background temperature at redshifts from 0 to 1.[1,2] Some additional background temperature estimates exist at redshifts from 1.8 to 3.3 based on molecular and atomic line excitation temperatures in quasar absorption line systems, but are model dependent.[3] To date, no deviations from the expected (1+$z$) scaling behavior of the microwave background temperature have been seen,[4] but the measurements have not extended deeply into the matter-dominated era of the universe at redshifts $z$>3.3. Here we report the detection of sub-millimeter line absorption from the water molecule against the cosmic microwave background at $z$=6.34 in a massive starburst galaxy, corresponding to a lookback time of 12.8 Gyr.[5] Radiative pumping of the upper level of the ground-state ortho-$H_2O(1_{10}$-$1_{01})$ line due to starburst activity in the dusty galaxy HFLS3 results in a cooling to below the redshifted microwave background temperature, after the transition is initially excited by the microwave background. The strength of this effect implies a microwave background temperature of 16.4-30.2 K (1σ range) at $z$=6.34, which is consistent with a background temperature increase with redshift as expected from the standard ΛCDM cosmology.[4]**

We used the Northern Extended Millimeter Array (NOEMA) to obtain a sensitive scan across the 3-millimeter atmospheric window toward the z=6.34 massive dusty starburst galaxy HFLS3 (also known as 1HERMES S350 J170647.8+584623; see Methods).[5] These observations reveal a broad range of emission features dominated by the CO, $H_2O$, and $H_2O^+$ molecules and atomic carbon, on top of thermal dust continuum emission that is rising with frequency consistent with a dust temperature of $T_{\rm dust}=63.3^{+5.4}_{-5.8}$ K (Figure 1).[5,6] The spectrum also shows a deep absorption feature due to the ortho-$H_2O(1_{10}$-$1_{01})$ ground state transition at rest-frame 538 μm (observed at 3.95 mm, or 75.9 GHz). This absorption is about two times stronger than the continuum emission from the starburst at the same frequency (Figure 2). For this effect to occur, a significant population of the ortho-$H_2O$ $1_{10}$ level (which lies 26.7 K above the $1_{01}$ ground state) has to be excited by cosmic microwave background (CMB) photons as a basis for pumping of this level by the starburst infrared radiation field (see Extended Data Figure 1). The effect becomes observable toward HFLS3 because of the warm CMB at this redshift, which is predicted to be $T_{\rm CMB}$=20.0 K at z=6.34 based on standard ΛCDM cosmology (where $T_{\rm CMB}(z) = T_{\rm CMB}(z=0)*(1+z)^{(1-\beta)}$, $T_{\rm CMB}(z=0)$=2.72548+/-0.00057 K,[7] and the power-law index $\beta$=0). The absorption of photons from the CMB radiation field significantly populates the $H_2O$ $1_{10}$ level. The intense infrared radiation field from the starburst then preferentially de-populates the $1_{10}$ level through radiative pumping, resulting in a deficit in the $1_{10}$ level compared to $1_{01}$ relative to a thermal distribution. In combination, these two processes result in an excitation temperature $T_{\rm ex}$ of the $H_2O(1_{10}$-$1_{01})$ line that is lower than $T_{\rm CMB}$, such that the line becomes observable in absorption against the CMB. Since the effect depends on the strength of the CMB radiation field, it can be used to measure $T_{CMB}$ for galaxies that have well-measured dust spectral energy distributions and dust continuum sizes, as is the case for HFLS3.

To understand the effect, we have calculated a series of spherically-symmetric RADEX[8] models over a wide range of $H_2O$ column densities, assuming purely radiative excitation (Figures 2 and 3; see Methods for additional details). Exposing a cold, $H_2O$-bearing region associated with HFLS3 to the blackbody CMB radiation field at $T_{CMB}(z=6.34)$, the models suggest that 77.2% of the molecules will be in the $1_{01}$ ground state and 20.3% will be in the upper $1_{10}$ state, and all $H_2O$ transitions have an excitation temperature $T_{ex}$ equal to $T_{CMB}$. As a result of this zero temperature contrast, no $H_2O$ emission or absorption would be observable, despite the fact that the $H_2O$ rotational ladder is excited by the CMB radiation. However, this picture changes when the same region is also exposed to the infrared radiation field emitted by the starburst nucleus of HFLS3, as the latter does not follow a single black-body radiation pattern. Indeed, the infrared spectral energy distribution of HFLS3 reaches its peak intensity at $73.3^{+1.6}_{-1.3}$ μm and can be approximated by a greybody with a Rayleigh-Jeans slope of $\beta_{IR}=1.94^{+0.07}_{-0.09}$. This is due to the presence of dust at multiple temperatures, and an increasing dust optical depth towards shorter wavelengths.[5,6] In this case, the level populations of the $1_{01}$ and $1_{10}$ states will deviate from the single-temperature thermal equilibrium population and change to 68.0% and 14.6%, respectively, for the solution shown in Figure 2, resulting in an excitation temperature $T_{ex}$ of only 17.4 K for this transition. Due to the $\Delta J=1$ selection rule for photon emission/absorption, only three ortho-$H_2O$ transitions contribute to the modification of populations in the $1_{01}$ and $1_{10}$ levels, namely the 538 μm $1_{10}$-$1_{01}$ and 180 μm $2_{12}$-$1_{01}$ transitions affecting the former, and the 108 μm $2_{21}$-$1_{10}$ transition affecting the latter (see Fig. 2; the $2_{21}$-$1_{01}$ transition is forbidden). The over-proportional de-population of the $1_{10}$ level occurs since the $H_2O(2_{21}$-$1_{10})$ transition at 108 μm dominates the modification of the level population.[9,10] This transition lies near the peak of the dust spectral energy distribution, where the dust emission has a higher optical depth than at 538 μm where the $H_2O(1_{10}$-$1_{01})$ transition occurs. The rise in dust optical depth with wavelength leads to an increased availability of 108 μm photons relative to 538 μm compared to the thermal equilibrium case of a single black-body radiation field. This implies that the $2_{21}$-$1_{10}$ transition at 108 μm is exposed to a more intense infrared radiation field than the $1_{10}$-$1_{01}$ transition at 538 μm. For the infrared radiation field of HFLS3, the models thus suggest that both $H_2O$ transitions should be found in absorption, but that the $2_{21}$-$1_{10}$ transition (which is not covered by our observations) will contribute a larger fraction to the $1_{10}$ level de-population than found for the thermal equilibrium case. The CMB photons, on the other hand, only provide the base population expected for the thermal blackbody equilibrium case. This causes a reduced excitation temperature in the $1_{10}$-$1_{01}$ transition compared to the thermal equilibrium case, unless the $1_{01}$ level is even more substantially de-populated due to the $H_2O(2_{12}$-$1_{01})$ transition at 180 μm. This, however, does not occur, since our models show that this transition is expected to be seen in emission (at a line strength ~3-5 times below a previously reported upper limit[5]). This is because the upper level population of the $2_{12}$-$1_{01}$ transition is also affected by the population of the $2_{12}$ level through the $2_{21}$-$2_{12}$ line, which appears in emission due to the pumping of its upper $2_{21}$ state by the $2_{21}$-$1_{10}$ absorption line. As such, the models suggest a net deficit in the upper level population of the $1_{10}$-$1_{01}$ transition compared to the thermal equilibrium case, which drives the $T_{ex}$ of the line to end up below the CMB temperature.

The RADEX models yield $T_{ex}$ = 17.4 K due to this level population modification. To translate model-predicted temperature differences into an absorption line flux that can be compared to the observations, the size of the emitting/absorbing region needs to be known. Based on the NOEMA observations at rest-frame 122 μm (Extended Data Fig. 2), we estimate the dust continuum size of the emitting region at 108 μm (i.e., the wavelength of the pumping transition) of HFLS3 to be $r_{108\mu m}$=1.62 +/- 0.45 kpc. Within the uncertainties of the size estimate, the RADEX models suggest that the strength of the observed $H_2O$ absorption can be reproduced over about two orders of magnitude in $H_2O$ column density, with a lower limit of $\sim10^{16}$ $cm^{-2}$. The minimum covering fraction of the dust continuum is ~60% when conservatively leaving $T_{CMB}$ as a free parameter (100% is assumed for the grid shown in Fig. 3a). The upper limit for the $H_2$ column density implied by the gas mass of HFLS3 of (1.04 +/- 0.09) x $10^{11}$ $M_{sun}$[5] provides a lower limit to the gas phase $[H_2O]/[H_2]$ abundance of >2 x $10^{-7}$, which falls within the range of $10^{-9}$-$10^{-5}$ found for nearby starbursts.[11] The small difference $\Delta T = T_{ex} - T_{CMB}$=-2.6 K thus is sufficient to explain the observed strength of the $H_2O(1_{10}-1_{01})$ absorption line toward the CMB in HFLS3 when assuming a layer of cold, diffuse $H_2O$-bearing gas with a high covering fraction in front of the warm dust continuum source associated with the $H_2O$ emission lines.

Since the absorption line is observed in contrast to the CMB, we can use the strength of the absorption line to obtain a measurement of $T_{CMB}$ at the redshift of HFLS3. The RADEX models suggest that, in order to detect the $H_2O(1_{10}-1_{01})$ line in absorption against the CMB, $T_{CMB}(z=6.34)$ must be >7-8 K independent of the model assumptions (see Methods). The observed strength of the signal suggests 16.4 K < $T_{CMB}(z=6.34)$ < 30.2 K (1σ, or 12.8 K < $T_{CMB}(z=6.34)$ < 34.0 K 2σ) for HFLS3 when treating $T_{dust}$, $\beta_{IR}$ , and the wavelength where the dust optical depth reaches unity as free fitting parameters for each dust continuum size sampled by the models. This explains why the effect has not been previously seen. $T_{CMB}$ must be sufficiently high to satisfy the requirement of a significant $H_2O$ $1_{10}$ level population due to the CMB, such that a de-population by the infrared radiation field of the starburst will lead to a sufficiently significant decrement to be observable in absorption against the CMB. This limits observability to z>4.5 for dust spectral energy distribution shapes and dust continuum sizes of star-forming galaxies like HFLS3 (Figure 3c), where only few spectra at rest-frame 538 µm with sufficient signal-to-noise ratio to detect the effect exist. This differs from molecules like $H_2CO$, for which absorption against the CMB has been predicted to occur at any redshift up to present day,[12] but for which no detections at high redshift currently exist.[13] For starbursts with dust as warm as HFLS3, its relative strength is expected to continue to increase with redshift all the way up to z~7-8, and to remain observable back to the earliest epochs when such galaxies existed.

The thermal Sunyaev-Zel'dovich (SZ) effect entails the modification of the thermal distribution of CMB photons by Thomson scattering off electrons at high temperatures in the intergalactic medium of galaxy (proto-) clusters. The effect observed here requires the CMB photons to excite the $H_2O$ rotational ladder to create a thermal distribution of the lower energy levels, which then is modified through the absorption of far-infrared photons from the starburst radiation field permeating the interstellar medium. In both cases, the CMB is responsible for establishing unperturbed thermal distributions (of photons and $H_2O$ excitation, respectively), which then get modified by local conditions. The SZ effect is a broadband modification of the thermal distribution of CMB photons via scattering, with an expected signal strength (relative to the CMB) that is independent of redshift. In contrast, the $H_2O$ absorption signal described herein is a narrow band (spectral line) absorption process of the CMB photons, catalyzed (in part) by the CMB itself, with an absorption line strength that increases with redshift due to the increasing temperature of the CMB, relative to the fixed excitation temperature of the $H_2O(1_{10}-1_{01})$ transition.

Standard ΛCDM cosmologies predict a linear increase of $T_{CMB}$ with $(1 + z)$. However, there are hypothetical physical mechanisms that could lead to departures from this linear behavior, including the evolution of physical constants,[14] decaying dark energy models,[15] and axion-photon-like coupling processes.[16,17] Direct measurements of $T_{CMB}$ thus are a crucial test of cosmology, but they currently are limited to $z$<1, due to the lack of sufficiently precise measurements of the thermal SZ effect in galaxy clusters at higher redshifts (Figure 4; see Methods for further details). A limited sample of additional constraints exists at $z$=1.8-3.3 based on measurements of $T_{ex}$ for the ultraviolet transitions of CO, [CI], and [CII] in absorption line systems along the lines of sight to quasars. These lines are not directly observed in contrast to the CMB, and they use the $T_{ex}$ of these lines as a proxy of $T_{CMB}$, such that the resulting $T_{CMB}$ estimates are subject to model-dependent excitation corrections.[3,17-28] As an example, for the CO molecule, $T_{ex}$ typically already exceeds $T_{CMB}$ in the diffuse interstellar medium in the Milky Way due to collisional excitation, showing a rising excess with increasing CO optical depth due to photon trapping.[18] In contrast, our models suggest that collisional excitation of $H_2O$ becomes important only at very high densities, such that $H_2O$-based measurements are likely only minimally affected by this effect. The $H_2O$ absorption against the CMB at $z$=6.34 reported here thus provides the most direct constraint on $T_{CMB}$ currently available at $z$>1. Indeed, the existence of this effect on its own directly implies that the CMB is warmer than at low redshift, because $T_{CMB}$ must be sufficiently high to significantly excite the $H_2O$ $1_{10}$ level, which lies 26.7 K above ground, as a basis for the observed decrement due to de-population of this level by the starburst radiation field. A combined fit to the available data (Figure 4) is consistent with the redshift scaling expected from ΛCDM. Fitting for the adiabatic index $\gamma$ in the equation of state between pressure $P$ and energy density $\rho$ for the sum of baryonic and dark matter and radiation, i.e., $P_{rm} = (\gamma\text{-}1)\ \rho_{rm}$ with a standard formalism (see Methods), we find $\gamma$=$1.328^{+0.008}_{-0.007}$, which agrees with the standard value of $\gamma$=4/3 expected in ΛCDM. At the same time, we find an effective dark energy equation of state parameter $w_{eff}$=$P_{de}/\rho_{de}$ =$-1.011^{+0.018}_{-0.017}$, which is consistent with the $w$=-1 expectation for a dark energy density that does not evolve with time.

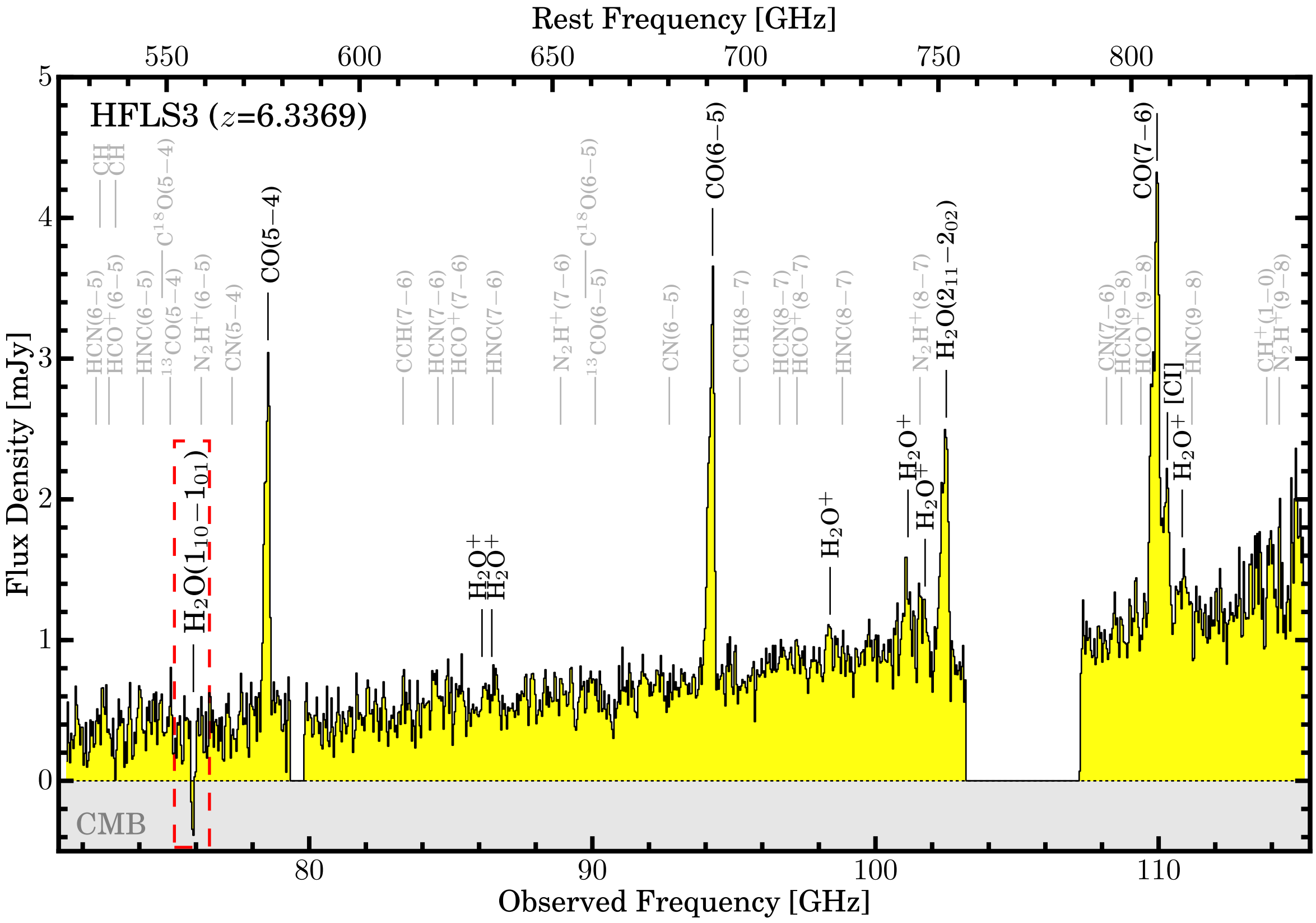

**Figure 1: Broad-band 3mm spectroscopy of the starburst galaxy HFLS3 at a redshift of 6.34 with NOEMA.** Black/yellow histogram, NOEMA spectroscopy data, binned to 40 MHz (158 $kms^{-1}$ at 75.9 GHz) spectral resolution. Expected frequencies of molecular and atomic lines at the redshift of HFLS3 are indicated, with the dominant species labeled in black. The dashed red box indicates the frequency range of the ortho-$H_2O(1_{10}-1_{01})$ line, which is detected in absorption against the CMB.

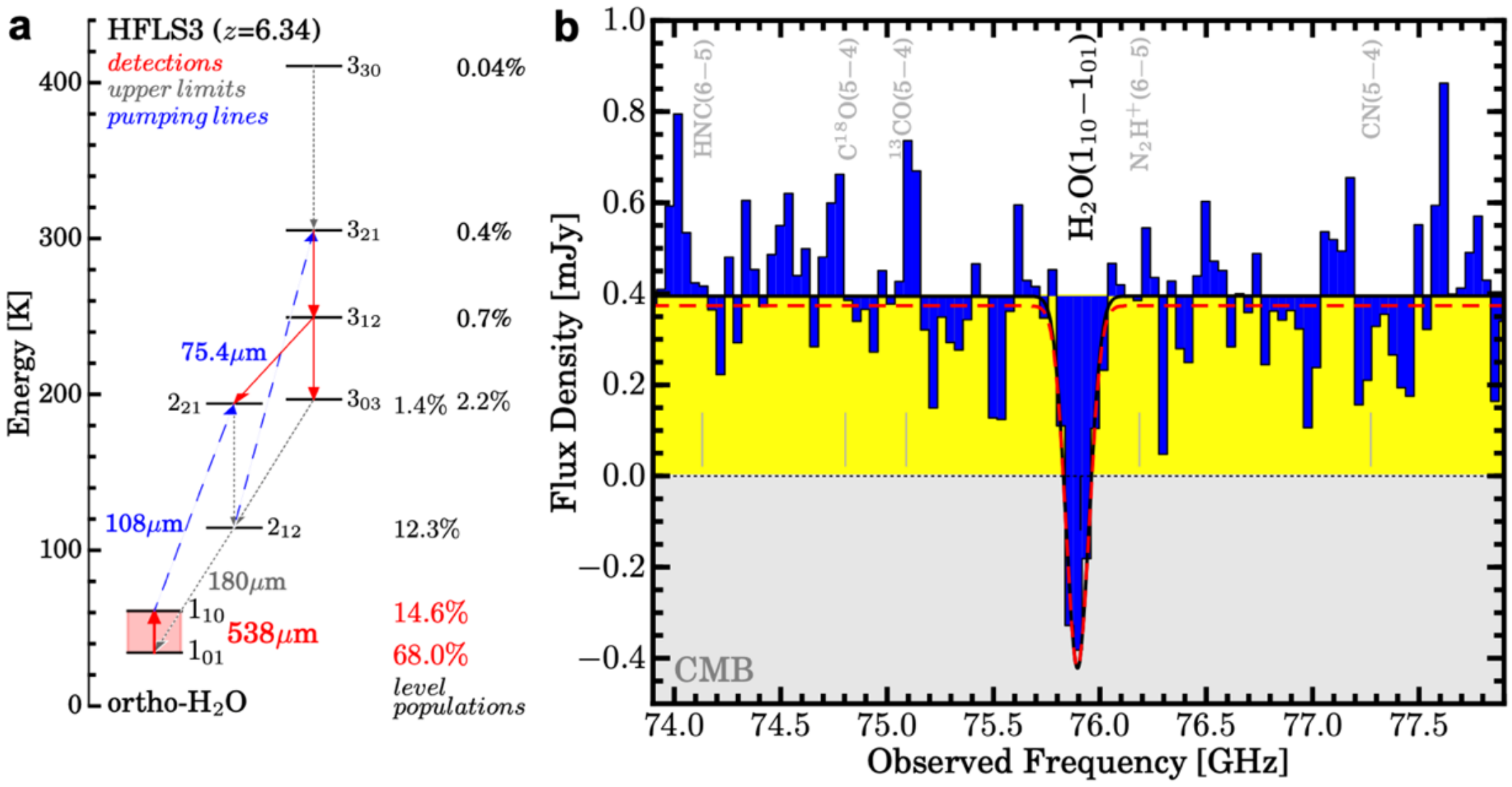

**Figure 2: $H_2O$ line and continuum properties and modeling of HFLS3**. Ortho-$H_2O$ energy level structure (**a,** solid arrows are detected transitions, dotted lines are upper limits, and dashed arrows are pumping transitions, with observed and model-predicted absorption/emission lines indicated as upward/downward arrows; percentages are the level populations in the model), and zoom-in on the $H_2O$ line at the same spectral resolution as in Fig. 1 to show that the line absorbs into the CMB (**b,** blue shading added for emphasis). The black curve is a fit to the spectrum. The red dashed curve is the best-fit radiative transfer model.

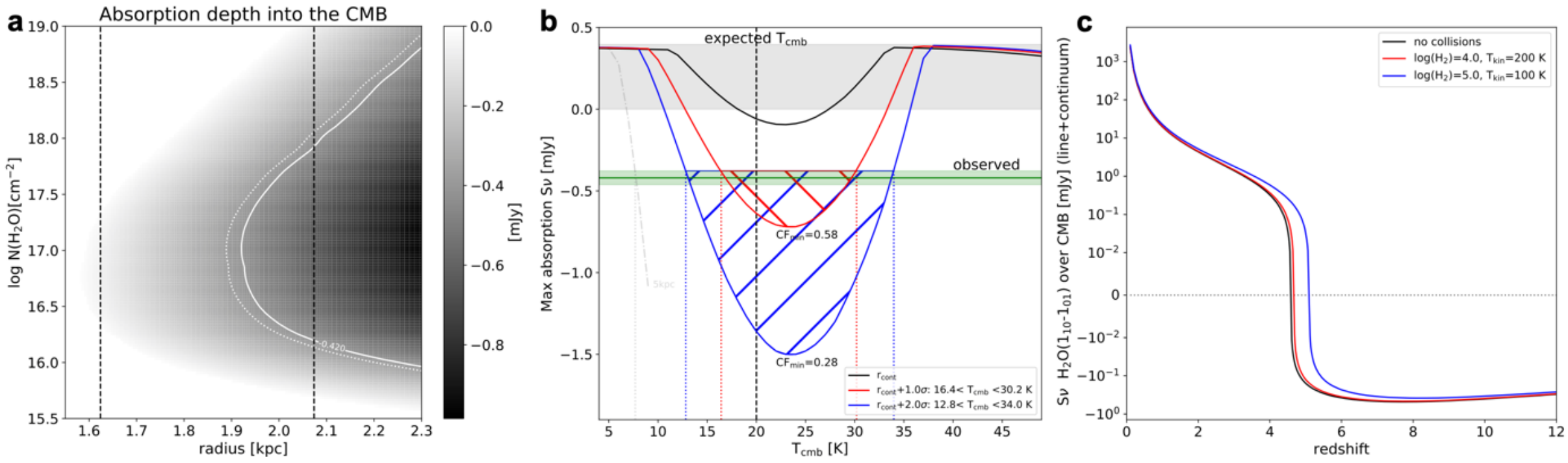

**Figure 3: Radiative transfer models for HFLS3, and constraints on the CMB temperature. a,** Model grid for the predicted line absorption strength for $T_{CMB}(z=6.34)=20.0$ K (grayscale) as a function of $H_2O$ column density ($y$-axis) and radius of the dust emission region at 108 µm ($x$-axis). The white curve shows the parameter space allowed by the measurement (solid) and -1-σ r.m.s. uncertainty region (dotted). The dashed black lines show the measured continuum size (left) and +1-σ r.m.s. uncertainty region (right). The overlapping region between within the white boundary (i.e., minimum allowed absorption strength) and the size measurement (i.e., minimum required emitting area at 100% covering fraction) is the allowed parameter space for the absorption strength within 1-σ r.m.s. The minimum in the required radius at $N(H_2O)\sim10^{17}$ cm$^{-2}$ is due to a minimum in $T_{ex}$ in the models. **b,** Constraints on $T_{CMB}$ for the observed absorption strength (green line and shaded region) at the minimum size compatible with the observations (**a**), based on the same models (red/blue shaded regions are the allowed ranges within the source radius +1-σ/+2-σ r.m.s.). The source radius at face value (black line), as well as the -1-σ and -2-σ r.m.s. regions (not shown) are ruled out by the observations. The minimum filling factor of the dust emission region $CF_{min}$ is indicated for the +1-σ and +2-σ r.m.s. regions. The dashed gray line shows a model assuming a continuum radius of 5 kpc, which provides a conservative lower limit on $T_{CMB}$. **c,** Observability of the $H_2O$ absorption as a function of redshift for three solutions allowed by the data without and with collisional excitation. The effect becomes observable at $z\sim4.5$ and remains visible at similar strength out to $z>12$. The lower-redshift limit is higher in cases where collisional excitation is important, but the impact is minor below $n(H_2)=10^5$ cm$^{-3}$.

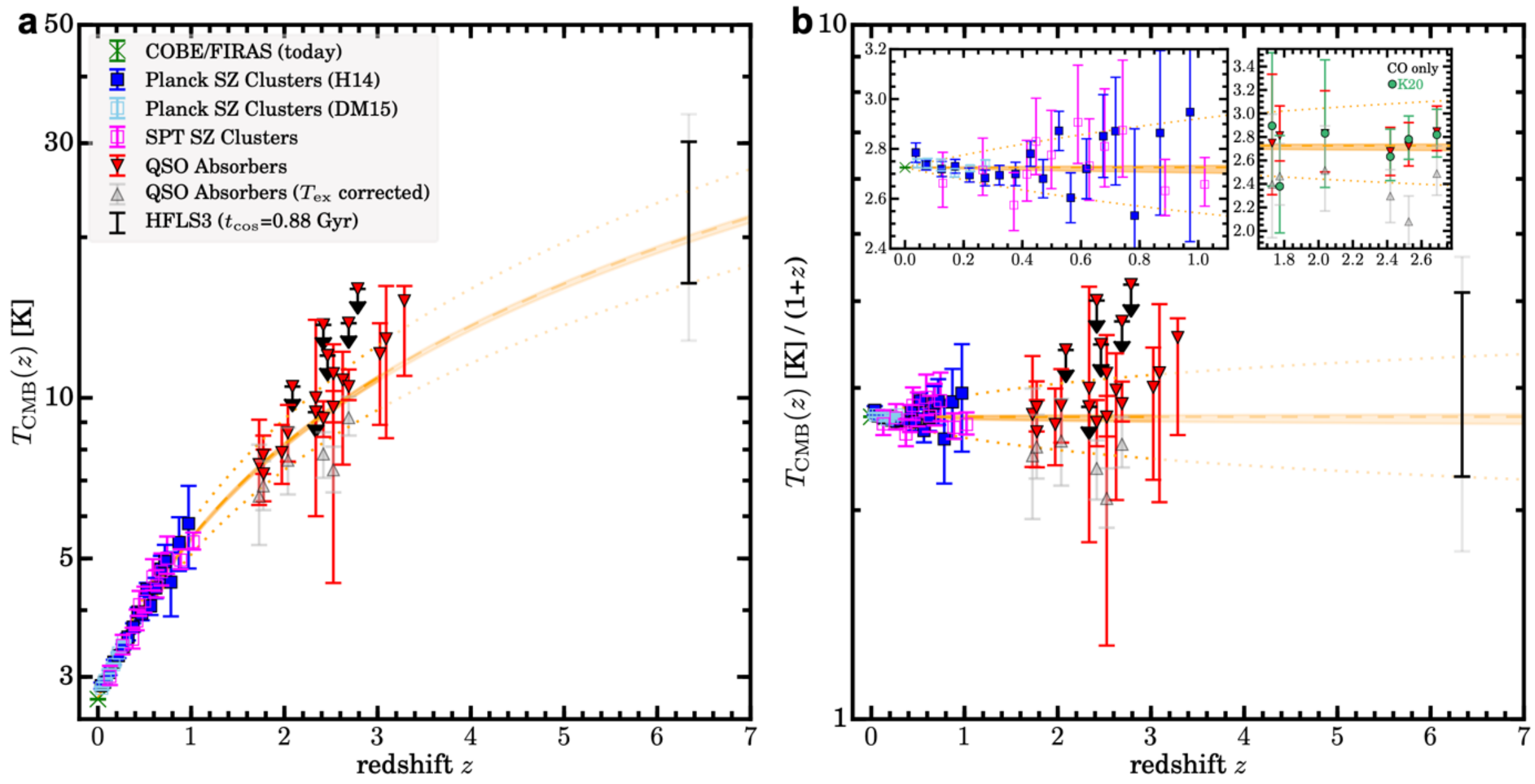

**Figure 4: Measurements of the CMB temperature as a function of redshift.**[3,17,19-28] **a,** 1-σ (black) and 2-σ r.m.s. (gray) uncertainties are shown for HFLS3, and 1-σ r.m.s. uncertainties elsewhere**. b,** Same data, but dividing out the (1+*z*) redshift scaling of the CMB expected from ΛCDM. Previous direct measurements are from CMB mapping at *z*=0 and Sunyaev-Zel'dovich effect measurements of galaxy clusters in contrast to the CMB out to *z*~1 (left zoom-in panel in **b**). Additional measurements are from UV absorption systems along the lines of sights to quasars out to *z*~3. The downward (upward) triangles are not corrected (corrected) for the contribution of collisional excitation in the diffuse ISM to the excitation temperature $T_{ex}$ of the tracer (right zoom-in panel in **b**; green dots show an alternative proposed correction[3]).[21,27] The separation of these pairs of points for the same sources is an indication of the systematic uncertainties on top of the statistical uncertainties indicated by the error bars. The $H_2O$-based measurement of HFLS3, like those up to *z*~1, is in contrast to the CMB, but as a line measurement, it is more precise in redshift. It is not subject to the same uncertainties in $T_{ex}$ as the intermediate redshift measurements, since collisions can only decrease (rather than boost) the resulting absorption strength into the CMB for the $H_2O$-based measurement. They are also unlikely to play an important role due to the high density required to collisionally excite the relevant $H_2O$ lines. Ignoring collisions results in the most conservative estimate of $T_{CMB}$ for HFLS3. The orange shaded region shows a $T_{CMB}=T_{CMB}(z=0)*(1+z)^{1-\beta}$ fit to the data in Extended Data Table 1 and its uncertainty (where $T_{CMB}(z=0)$=2.72548+/-0.00057 K,[7] and $\beta=(3.4^{+8.1}_{-7.3})$ x $10^{-3}$), the orange dashed line indicates the $\beta$=0 case corresponding to the standard cosmology, and the dotted lines indicate a +/-10% deviation in 1-$\beta$.

**Methods:**

**NOEMA observations.** The target was observed in the 3 mm wavelength band 1 (rest-frame 400 μm) with NOEMA as part of project S20DA (PIs: D. Riechers, F. Walter). Three partially overlapping spectral setups were observed under good weather conditions between 2020 July 26 and August 25 with 10 antennas in the most compact D configuration, using a bandwidth of 7.7 GHz (dual polarization) at 2 MHz spectral resolution per sideband. We also included previously-published[5] observations between 2012 February 06 and May 31 in the A and D configuration tuned to 110.128 and 113.819 GHz, respectively, and previously unpublished observations between 2012 June 01 and 04 and on 2017 July 10 in the D configuration tuned to 78.544 and 101.819 GHz taken as part of projects V0BD, W058, and S17CC (PI: D. A. Riechers), all using 3.6 GHz bandwidth (dual polarization), yielding 21 observing runs total. Nearby radio quasars were used for complex gain, bandpass, and absolute flux calibration. The target was also observed in the 0.87 mm wavelength band 4 (rest-frame 122 μm) with NOEMA as part of project X0CC (PI: D. Riechers). Observations were carried out during 3 observing runs with 6 antennas in the A and C configurations under good weather conditions between 2013 December 4 and 2015 March 12, with the band 4 receivers tuned to 335.5 GHz, and using a bandwidth of 3.6 GHz (dual polarization). Nearby radio quasars were used for complex gain, bandpass, and absolute flux calibration. The GILDAS package was used for data calibration and imaging. All 3 mm data were combined to a single visibility cube before imaging. Imaging was carried out with natural baseline weighting. The band 4 data were also imaged with Briggs robust weighting to increase the spatial resolution. A map of the continuum emission at the frequency of the $H_2O$ line was created by averaging the visibility data over a bandwidth of 2.04 GHz centered on the line. This range was chosen to avoid other lines in the bandpass. Continuum emission was subtracted from the $H_2O$ line cube in the visibility plane. Moment 0 images of the line absorption were created before and after continuum subtraction by integrating the signal over a bandwidth of 100 MHz, corresponding to 395 km $s^{-1}$. The resulting rms noise levels are provided in Extended Data Figure 2. We also make use of previously-published[5] rest-frame 158 μm NOEMA data, which were adopted without further modification.

**Line and Continuum Parameters.** The flux of the $H_2O(1_{10}\text{-}1_{01})$ line was extracted by simultaneous Gaussian fitting of the line and continuum emission (including a linear term for the continuum) in the one-dimensional spectrum shown in Figure 1, which was extracted from the image cube. The source is unresolved at the frequency of the $H_2O(1_{10}\text{-}1_{01})$ line, such that the main uncertainties are due to the slope of the continuum emission and the appropriate fitting of other nearby lines, in particular CO(5-4). The uncertainties in these parameters are part of the quoted uncertainties. We find a line peak flux of -818 +/- 145 μJy at a line FWHM of 507 +/- 111 km $s^{-1}$, centered at a frequency of 75.8948 GHz (+/- 46 km $s^{-1}$; the calibration uncertainties on the line FWHM and center frequency are negligible, and that on the line peak flux is <10%, i.e., minor compared to the measurement uncertainty). Given the rest frequency of the line of 556.9359877 GHz, this corresponds to a redshift of 6.3383, which is consistent with the systemic redshift of HFLS3 (z=6.3335 and 6.3427 with uncertainties of +/-14 and +/-54 km $s^{-1}$ at Gaussian FWHM of 243 +/- 39 and 760 +/- 152 km $s^{-1}$, respectively, for the two velocity components detected in the 158 μm [CII] line).[5] For comparison, the $H_2O$ $2_{02}\text{-}1_{11}$ and $2_{11}\text{-}2_{02}$ emission lines in HFLS3 have FWHM of 805 +/- 129 and 927 +/- 330 km $s^{-1}$, respectively,[5] i.e., only marginally broader than the $1_{10}\text{-}1_{01}$ line at the current measurement uncertainties. The continuum flux at the line frequency is 396 +/- 15 μJy, corresponding to 48% +/- 9% of the absorption line flux (the relative flux calibration uncertainty between the line and continuum emission is negligible). We also measured the 335.5 GHz continuum flux by two-dimensional fitting to the continuum emission in the visibility plane. We find a flux of 33.9 +/- 1.1 mJy, which agrees with previous lower-resolution observations at the same wavelength.[5] The major (minor) axis FWHM diameter of the source is 0.617 +/- 0.074 arcsec (0.37 +/- 0.20 arcsec). This yields the physical source size quoted in the main text at the redshift of HFLS3.

**Brightness Temperature Contrast.** The $H_2O(1_{10}\text{-}1_{01})$ line leads to a decrement in continuum photons from the starburst, and as such, is observed as a lack of continuum emission at its frequency at the position of the starburst. It thus appears as negative flux in an image where starburst continuum emission has been subtracted.

In addition, (sub-) millimeter wavelength interferometric images reveal structure against a flat sky background defined by the large-scale CMB surface brightness, which the interferometer does not detect itself due to its limited spatial sampling. Thus, the fraction of the signal due to the decrement in CMB photons at the position of the starburst appears as negative flux without subtracting any further signal, but it also corresponds to a lack of continuum emission at the line frequency in practice. Since the mere presence of an absorption line signal stronger than the measured continuum emission implies absorption against the CMB, this interpretation is not limited by uncertainties in the galaxy continuum flux or uncertainties in the absolute flux calibration.

**Line Excitation Modeling.** RADEX is a radiative transfer program to analyze interstellar line spectra by calculating the intensities of atomic and molecular lines, assuming statistical equilibrium and considering collisional and radiative processes, as well as radiation from background sources. Optical depth effects are treated with an escape probability method.[8] Studies of nearby star-forming galaxies show that the observed absorption strengths of the ground-state $H_2O$ and $H_2O^+$ transitions are due to cooler gas that is located in front of, and irradiated by a warmer background source that is emitting the infrared continuum light that also excites the higher-level $H_2O$ emission lines.[11,29] We thus adopt the same geometry for the modeling in this work, which is adequately treated within RADEX (i.e., treating the dust continuum plus the CMB as background fields for the absorbing material).[8] The dust continuum emission is modeled as a greybody with treating $T_{\rm dust}$, $\beta_{\rm IR}$, and the wavelength where the dust optical depth reaches unity as free fitting parameters for each dust continuum size and $T_{\rm CMB}$ sampled by the models. The observed spectral energy distribution of HFLS3, including all literature[5] photometry and the measurements presented in this work, is then treated as the contrast between the dust continuum and CMB background fields, such that the resulting fit parameters for the dust continuum source change with $T_{\rm CMB}$ in a self-consistent manner. In the RADEX models, we derive the $H_2O$ peak absorption depth into the CMB. We then multiply the best matching peak absorption depth found by RADEX with a Gaussian matched to the fitted line centroid and line width obtained from the observed line profile in Figure 2 to determine the model line profile. In this approach, the shallower absorption in the line wings either corresponds to a lower filling factor of the $H_2O$ layer at the corresponding velocities, or to lower $H_2O$ column densities. While collisions of $H_2O$ molecules with $H_2$ are another mechanism that can modify the level populations especially at very high gas densities (which is an important mechanism for the cooling of low-excitation temperature transitions of molecules like $H_2CO$ to below $T_{\rm CMB}$),[12,30] the RADEX models show that they do not affect our findings (see Figure 3c). We thus adopt models with essentially no collisions by assuming a very low gas density of $n(H_2) = 10$ cm$^{-3}$. We then compare our findings to those obtained when adopting conditions that are similar to those found in local starburst galaxies,[11] and to those found for high-density environments with $n(H_2) > 10^5$ cm$^{-3}$. The cross sections for collisions out of the $1_{01}$ level are always larger than those out of the $1_{10}$ level, independent of the collision partner and the temperature at which the collisions take place.[31-33] Thus, collisions cannot be responsible for an over-proportional de-population of the $1_{10}$ level relative to the $1_{01}$ ground state, and the net effect of including collisions is a decrease in the absorption depth into the CMB by reducing the $T_{\rm CMB}$-$T_{\rm ex}$ temperature difference at very high gas densities compared to cases without collisions. For reference, the effect of collisions on the determination of $T_{\rm CMB}$ is negligible for the typical conditions found in local starbursts (i.e., $n(H_2) \sim 10^4$ cm$^{-3}$; $T_{\rm kin}$=20-180 K),[11] and only starts to have an impact for very high densities $n(H_2) > 10^5$ cm$^{-3}$. For a given continuum source size, the constraints on $T_{\rm CMB}$ thus would be tighter (i.e., would more quickly become inconsistent with the observations) for the high-density case than for the case without collisions, such that the latter approach is more conservative (see Figure 3b). The overall impact of collisional excitation thus would be more stringent requirements on the source size, covering fraction, and water column, such that their inclusion would only further strengthen our conclusions. We note that this is the opposite effect than in the case of the studies of UV lines,[3,17,19-28] where neglecting collisional excitation results in less conservative constraints on $T_{\rm CMB}$. If we were to assume that the $H_2O$ absorption were to emerge from within the infrared continuum-emitting region, a larger source size would likely be required to obtain the same absorption line strength due to a reduced effective radiation field strength from the starburst. Previous modeling attempts of nearby galaxies assuming such geometries have not been able to produce $H_2O(1_{10}$-$1_{01})$ line absorption on the scales necessary to explain the observations of HFLS3, which may indicate that even more complex assumptions would be required.[11] Thus, the resulting constraints would once again be less conservative, perhaps acting in a similar manner as the high-density case. Excluding both of these effects from the models thus leads to a maximally conservative estimate of $T_{CMB}$ and its uncertainties. Assuming a plane-parallel or similar geometry instead of a spherical geometry would only have a minor impact on our findings.[8] The models shown in Figure 3 assume a filling factor of unity, which is the most conservative possible assumption. A more clumpy geometry with a lower covering fraction remains possible for all $T_{CMB}$ values for which the predicted absorption strength exceeds the observed value (see shaded regions in Figure 3b). For reference, the minimum covering fractions consistent with the continuum size at the observed signal strength are shown for the different cases considered in Figure 3. The line absorption is also found to be optically thick, with an optical depth of $\tau_{H2O}$=21.1 for the solution shown in Figure 2b. To determine the redshift above which the effect becomes observable (Figure 3c), we fixed $r_{108\mu m}$, $T_{dust}$, $\beta_{IR}$, and $M_{dust}$ to the observed values, and the $H_2O$ column density to the value corresponding to the model spectrum. $H_2O$ line absorption into the dust continuum of HFLS3 would already become visible at $z$>2.9, but absorption into the CMB only becomes observable at $z$>4.5 (or higher for $H_2$ densities of >$10^5$ cm$^{-3}$). These values account for changes in the shape of the dust greybody spectrum (i.e., changes in the relative availability of 538 and 108 μm photons) due to changes in $T_{CMB}$ with redshift. To better quantify the impact of different modeling parameters, we have varied $T_{dust}$ and $\beta_{IR}$ beyond their previously estimated uncertainties (nominal reference values without considering variations in $T_{CMB}$ from the literature are $T_{dust}$=$63.3^{+5.4}_{-5.8}$ K and $\beta_{IR}$=$1.94^{+0.07}_{-0.09}$).[5,6] This is necessary, because both parameters are dependent on the varying $T_{CMB}$ in our models (and thus, are changing parameters in Figure 3b and c), such that their true uncertainties need to be re-evaluated. We independently varied $\beta_{IR}$ in the 1.6-2.4 range, and $T_{dust}$ in the +/-20 K range as functions of $T_{CMB}$ around the best-fit values. This shows that $\beta_{IR}$>2.0 and $T_{dust}$ lower by more than 10 K from the best fits yield very poor fits to the spectral energy distribution data, while $\Delta\beta_{IR}$>-0.1 below the best-fit value would require a larger continuum size than the measured $r_{108\mu m}$+1σ and thus are disfavored by the size constraint. Excluding these ranges, the extrema across this entire range would extend the uncertainty range in the predicted $T_{CMB}$ by only (-1.7; +5.4 K) and (-0.8; +4.4 K) for the $r_{108\mu m}$+1σ and $r_{108\mu m}$+2σ cases, respectively. For comparison, the difference between the +1σ and +2σ uncertainty ranges is (-3.6; +3.8 K). This shows that the impact of the uncertainties in the dust spectral energy distribution fitting parameters on those in $T_{CMB}$ are subdominant to those in the continuum size measurement. Conversely, we have studied the impact of changes in $T_{CMB}$ on the best-fit $T_{dust}$ and $\beta_{IR}$. For the values corresponding to $r_{108\mu m}$+1σ and $r_{108\mu m}$+2σ ranges, $T_{dust}$ typically changes by <0.5 K and $\beta_{IR}$ typically changes by <0.1-0.2 when varying the parameters independently. These changes are larger than the actual uncertainties, because the fit to the dust spectral energy distribution becomes increasingly poorer with these single-parameter variations. At the same time, these changes are subdominant to those induced by changes in dust continuum size within the +1σ and +2σ uncertainty ranges, which is consistent with our other findings.

**Other $H_2O$ transitions in HFLS3.** Five $H_2O$ lines were previously detected toward HFLS3 ($2_{02}$-$1_{11}$, $2_{11}$-$2_{02}$, $3_{12}$-$2_{21}$, $3_{12}$-$3_{03}$, and $3_{21}$-$3_{12}$), and two additional lines were tentatively detected ($4_{13}$-$4_{04}$ and $4_{22}$-$4_{13}$).[5] The $J_{up}$=3 transitions are due to ortho-$H_2O$, and all other transitions are due to para-$H_2O$. All of these transitions appear in emission. Given the high critical densities of these transitions, our RADEX models cannot reproduce the strength of these lines simultaneously with the observed ortho-$H_2O(1_{10}$-$1_{01})$ absorption strength, which suggests that they emerge from different gas components. For reference, to reproduce the strength of the $H_2O(2_{11}$-$2_{02})$ in Fig. 1 alone with collisional excitation, $n(H_2)$ = 2 x $10^7$ cm$^{-3}$ and $T_{kin}$=200 K would be required, but the $H_2O(1_{10}$-$1_{01})$ would no longer appear in absorption against the CMB if it were to emerge from the same gas component. This is consistent with the picture that the $H_2O$ absorption is due to a cold gas component along the line of sight to the warm gas that gives rise to the emission lines.[11] Observations of the para-$H_2O(1_{11}$-$0_{00})$ ground-state do not currently exist for HFLS3, but our models do not show this line in absorption toward the CMB.

**Origin of the lower and upper limits on $T_{\rm CMB}$.** Our models show that the lower limit on $T_{\rm CMB}$ at a given redshift based on the observed $H_2O$ absorption is due to the minimum "seed" level population due to the CMB blackbody radiation field. To determine a conservative lower limit, we have calculated models with continuum sizes up to $r_{108\mu m}$=5 kpc (see Fig. 3b), corresponding to a +7.5σ deviation from the observed continuum size, and recorded at which temperature such weakly-constrained models turn into absorption. We find that this results in a lower limit of $T_{\rm CMB}$>7-8 K, independent of the model assumptions. This finding alone does not explain the existence of an upper limit in Fig. 3b. For a given size of the dust continuum emission, an increase in $T_{\rm CMB}$ also requires an increase in $M_{\rm dust}$ to still reproduce the observed dust SED, which leads to an effective increase in the dust optical depth at a given wavelength. The result of a rising optical depth is that the greybody spectrum between 538 and 108 μm increasingly resembles a blackbody spectrum, and hence, a decrease in the $H_2O$ absorption against the CMB. This effect is responsible for the upper limit in allowed $T_{\rm CMB}$ for a given dust continuum size and absorption strength.

**Uncertainties of $T_{\rm CMB}$ Measurements.** The uncertainties shown for the literature data in Figure 4 are adopted from the literature sources without modification, and they typically represent the statistical uncertainties from the individual measurements or sample averages. Individual cluster measurements of the thermal SZ effect may be affected by dust associated with foreground galaxies or the Milky Way, the galaxy clusters, or background galaxies that may be amplified by gravitational lensing, uncertainties in the reconstruction of the Compton-$y$ parameter maps due to flux uncertainties, radio emission due to active galactic nuclei and/or relics, the kinetic and relativistic SZ effects, and general bandpass and calibration uncertainties.[17] Furthermore, uncertainties on the cluster geometry, and thus, line-of-sight travel distance of the CMB photons through the cluster, and on the temperature of the intra-cluster gas limit the precision of individual SZ measurements. Sample averages may also be affected by systematics in the stacking procedures. Individual data points deviate by up to at least two standard deviations from the trend, which may indicate residual uncertainties beyond the statistical error bars provided, such that the error bars shown in Figure 4 are underestimated. The main source of uncertainty for the ultraviolet absorption line-based measurements are due to the assumption of no collisional excitation, which is not taken into account in the statistical uncertainties shown in Figure 4. Attempts to take this effect into account appear to suggest significantly larger uncertainties than indicated by individual error bars (Figure 4).[27] To expand upon earlier estimates,[21] we have calculated RADEX models for typical $T_{\rm kin}$, $n$(H), and column densities found from [CI] measurements in the diffuse ISM,[34] which suggests that collisional excitation contributes to the predicted $T_{\rm ex}$ of the lower fine structure transition. While we show the original unmodified data, the ultraviolet-based measurements thus are subject to uncertainties due to model-dependent excitation corrections on top of the statistical uncertainties. Furthermore, the fine structure levels of tracers like the [CI] lines can be excited by ultraviolet excitation and following cascades. To constrain $T_{\rm CMB}$ based on these measurements, the kinetic temperature, particle density, and local ultraviolet radiation field must be known, and are typically determined based on other tracers than the species used to constrain $T_{\rm CMB}$. Also, some measurements are based on spectrally unresolved lines, which limits the precision of kinetic temperature measurements based on thermal broadening.[21] Due to these uncertainties, the ultraviolet absorption line-based measurements are likely consistent with the standard ΛCDM value, but they do not constitute a direct measurement of $T_{\rm CMB}$ without significant further assumptions. For reference, the median $T_{\rm CMB}/(1+z)$ estimate based on the [CI] measurements alone (excluding upper limits) is 3.07 K with a median absolute deviation of 0.09 K and a standard deviation of 0.31 K. Thus, the current sample median deviates from the ΛCDM value by about one standard deviation. A combination of the (uncorrected) CO, [CI], and [CII]-based measurements provides a median value of 2.84 K with a median absolute deviation of 0.15 K and a standard deviation of 0.25 K. This highlights the importance of the corrections discussed above and in the literature, and the value of measurements with systematic uncertainties that differ from this method to obtain a more complete picture. The main source of uncertainty of the $H_2O$-based measurements, beyond the caveats stated in the line excitation modeling section, are the statistical uncertainties on the source size, the lack of a direct measurement of the absorbing $H_2O$ column density, variations in the dust mass absorption coefficient, and the filling factor. Given the high metallicity suggested by other molecular line detections, the limitation to high filling factors due to the source size, and the constraint on the gas mass from dynamical mass measurements, the main source of uncertainty resides in the source size due to limited spatial resolution in the current data. As such, major improvements should be possible by obtaining higher, (sub-) kiloparsec resolution (i.e., <0.2”) imaging with the Atacama Large Millimeter/submillimeter Array (ALMA; for other targets) and planned upgrades to NOEMA, and in the future, with the Next Generation Very Large Array (ngVLA). Statistical uncertainties will also be significantly reduced by observing larger samples of massive star-forming galaxies over the entire redshift range where measurements are possible, closing the gap to SZ-based studies which are currently limited to $z<1$. The resulting improvement in precision will provide the constraints that are necessary to confirm or challenge the evolution of the CMB temperature with redshift predicted by standard cosmological models.

**Accessibility of the Line Signal.** The frequency range currently covered by NOEMA is 70.4-119.3, 127.0-182.9, and 196.1-276.0 GHz (with significantly reduced sensitivity above ~115 and ~180 GHz in the first two frequency ranges). ALMA covers the 84-500 GHz range with gaps at 116-125 and 373-385 GHz, with a future extension down to 65 GHz (with significantly reduced sensitivity below ~67 GHz). The ngVLA is envisioned to cover the 70-116 GHz range. Excluding regions of poor atmospheric transparency, the $H_2O(1_{10}$-$1_{01})$ line thus is observable in these frequency ranges at redshifts of $z$=0.1-0.4, 0.5-2.0, 2.1-3.4, and 3.8-6.9 in principle, but the detectability of the line in absorption against the CMB is likely limited to the $z$~4.5-6.9 range if the SED shape of HFLS3 is representative. At lower frequencies, the Karl G. Jansky Very Large Array, and in the future, ALMA and the ngVLA, also provide access to the <52 GHz range, such that the signal also becomes observable at $z>9.7$ in principle. In conclusion, the absorption of the ground-state $H_2O$ transition against the CMB identified here could be traced from the ground toward star-forming galaxies across most of the first ~1.5 billion years of cosmic history.

**Detectability of the Line Signal for different SED shapes.** To investigate if the effect is expected to be detectable for different galaxy populations, we have applied our modeling to the $z$=3.9 quasar APM 08279+5255, for which the dust SED is composed of a dominant 220 K dust component and a weaker 65 K dust component, contributing only 10%-15% to the far-infrared luminosity.[35] The models suggest that the line is expected to occur in emission, and that it would not be expected to be detectable in absorption at any redshift out to at least $z$=12 in galaxies with similar dust SEDs. Other far-infrared-luminous high-redshift active galactic nucleus host galaxies typically show a stronger relative contribution of their lower-temperature dust components, such that the effect may remain detectable in less extreme cases. For galaxies with lower dust temperatures than HFLS3, the effect may be present even at lower redshifts, but is typically expected to be weaker in general and to disappear at redshifts where $T_{\rm CMB}$ approaches their $T_{\rm dust}$. For a dust SED shape resembling the central region of the Milky Way but otherwise similar properties, the effect is expected to be reduced by more than two orders of magnitude at its redshift peak, and to become virtually unobservable at the redshift of HFLS3. Thus, dusty starburst galaxies appear to be some of the best environments to detect the effect.

**Derivation of equation of state parameters:** The determine the adiabatic index, we assume a standard Friedmann-Lemaitre-Robertson-Walker cosmology with zero curvature and a matter-radiation fluid that follows the standard adiabatic equation of state quoted in the main text. This would correspond to a redshift scaling $T_{\rm CMB}(z)=T_{\rm CMB}(z=0) * (1+z)^{3(\gamma-1)}$ in the presence of a dark energy density that does not scale with redshift. The dark energy density is parameterized to scale with a power law $(1+z)^m$, where $m$=0 corresponds to a cosmological constant. With standard assumptions, this yields a redshift scaling of $T_{\rm CMB}$:[15]

$$T_{CMB}(z) = T_{CMB}(z=0)(1+z)^{3(\gamma-1)}\left[\frac{(m-3\Omega_{m,0}) + m(1+z)^{(m-3)}(\Omega_{m,0}-1)}{(m-3)\Omega_{m,0}}\right]^{(\gamma-1)}$$

and an effective dark energy equation of state $P_{\rm de}=w_{\rm eff}\,\rho_{\rm de}$, where the effective equation of state parameter $w_{\rm eff}=(m/3)-1$. This fitting function is used here with a canonical value of $\Omega_{\rm m,0}=0.315$.[4] The uncertainty of $\Omega_{\rm m,0}$ is small compared to all other sources of uncertainty and hence neglected.

All data used in the fitting are provided in Extended Data Table 1.[36-46]

**Acknowledgments:** D.R. acknowledges support from the National Science Foundation under grant numbers AST-1614213 and AST-1910107. D.R. also acknowledges support from the Alexander von Humboldt Foundation through a Humboldt Research Fellowship for Experienced Researchers. This work is based on observations carried out under project numbers V0BD, W058, X0CC, S17CC, and S20DA with the IRAM NOEMA Interferometer. IRAM is supported by INSU/CNRS (France), MPG (Germany) and IGN (Spain).

**Author Contributions:** D.R. led the project and writing of the manuscript. A.W. produced the models used in this work. F.W. was a co-Principal Investigator of the main observing proposal and contributed to the interpretation of the results. R.N. calibrated the data. All authors have reviewed, discussed, and commented on the manuscript.

**Competing Interest Statement:** The authors declare that they have no competing financial interests.

**Materials & Correspondence:** Correspondence and requests for material should be addressed to Dominik Riechers (riechers@ph1.uni-koeln.de).

**Data availability:** The spectral line data and model generated and analyzed during this study as shown in Figure 2 are linked to this manuscript in spreadsheet form. Additional versions of the NOEMA datasets (visibilities, images, and spectra) are available from the corresponding author (D.R.) on reasonable request. All data are also available in the IRAM Science Data Archive (isda@iram.fr) under project IDs V0BD, W058, X0CC, S17CC, and S20DA.

**Code availability:** The RADEX code used for the modeling presented in this work and shown in Figure 3 is available at https://home.strw.leidenuniv.nl/~moldata/radex.html .

Reprints and permissions information is available at www.nature.com/reprints .

**Extended Data:**

A schematic explanation of the observed effect is provided in Extended Data Figure 1, and line/continuum maps and additional information on the source size in Extended Data Figure 2.

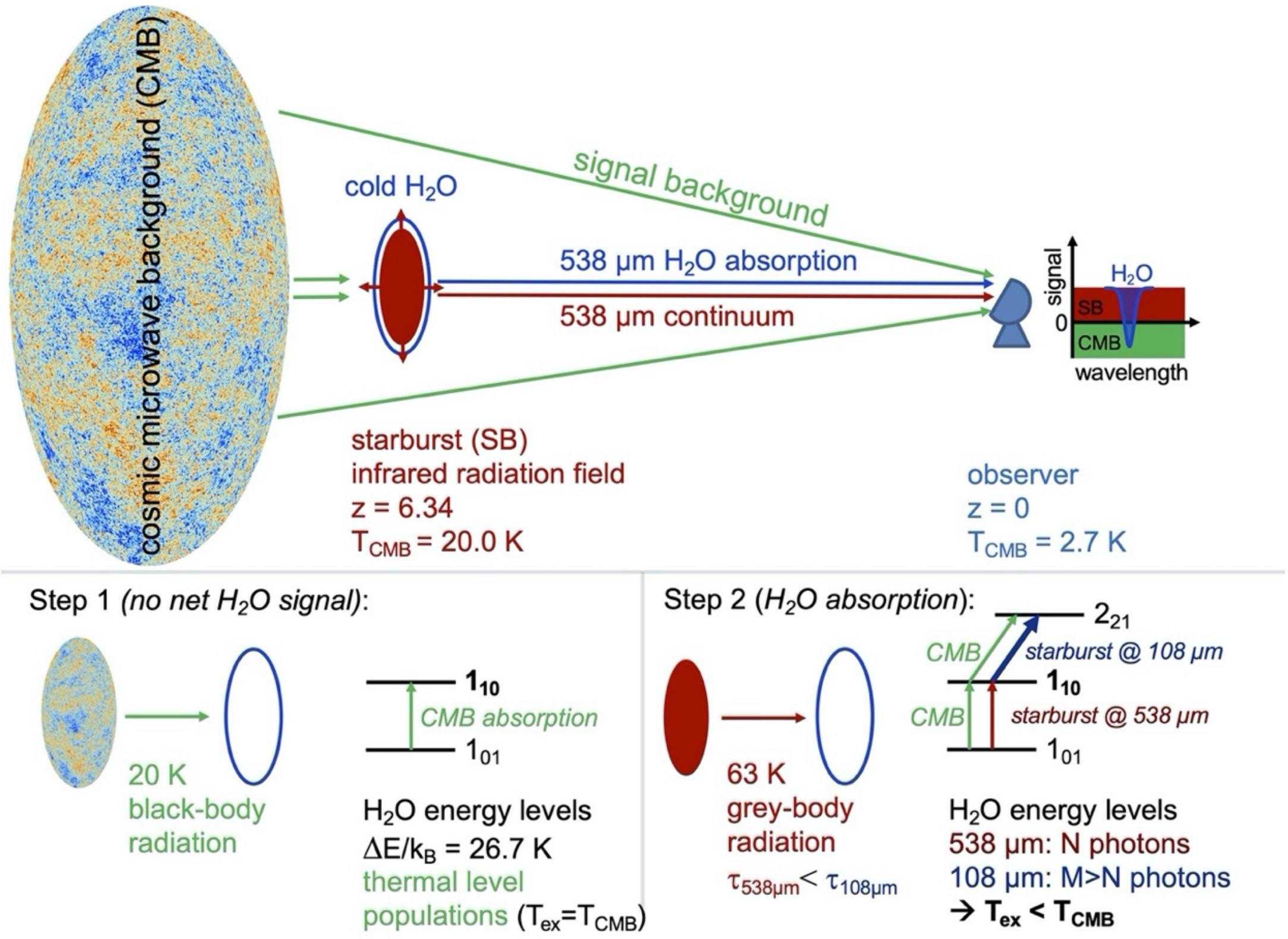

**Extended Data Figure 1: Combined Effect of CMB absorption and starburst radiation field on the strength of the $H_2O$ $1_{10}$-$1_{01}$ line in HFLS3.** Top: The cold $H_2O$ vapor is exposed to the CMB radiation field,[4] which has the shape of a black-body function ($T_{CMB}$=20.0 K at $z$=6.34), and the starburst infrared radiation field, which has the shape of a grey-body function ($T_{dust}$=63 K). NOEMA observed the signal in contrast to the CMB, and thus, detects only the dust emission from the starburst and the $H_2O$ line, but not the CMB itself (which thus fills the region below zero flux density as seen by the telescope). Bottom left: Since the energy level difference for the $H_2O$ $1_{10}$-$1_{01}$ line is only 26.7 K, there are sufficient CMB photons at $z$=6.34 to thermalize the level population between both levels, such that $T_{ex}$ is the same as $T_{CMB}$ in equilibrium. Thus, no $H_2O$ emission or absorption will be observed despite the presence of a "seed" population in the upper level. Bottom right: The radiation field of the starburst alters the level populations toward increased higher-level populations. Due to the grey-body shape of its spectral energy distribution, more photons are available at 108 µm to increase the $2_{21}$ level population from the $1_{10}$ state than there are 538 µm photons available to increase the $1_{10}$ level population from the $1_{01}$ state, relative to the "seed" population provided by the absorption of CMB photons. Thus, the relative population of the $1_{10}$ and $1_{01}$ levels is lower than in thermal equilibrium, such that the resulting $T_{ex}$ is lower than $T_{CMB}$. As a result, the $H_2O$ $1_{10}$-$1_{01}$ line is observed in absorption toward the CMB due to the negative temperature contrast – as observed toward HFLS3.

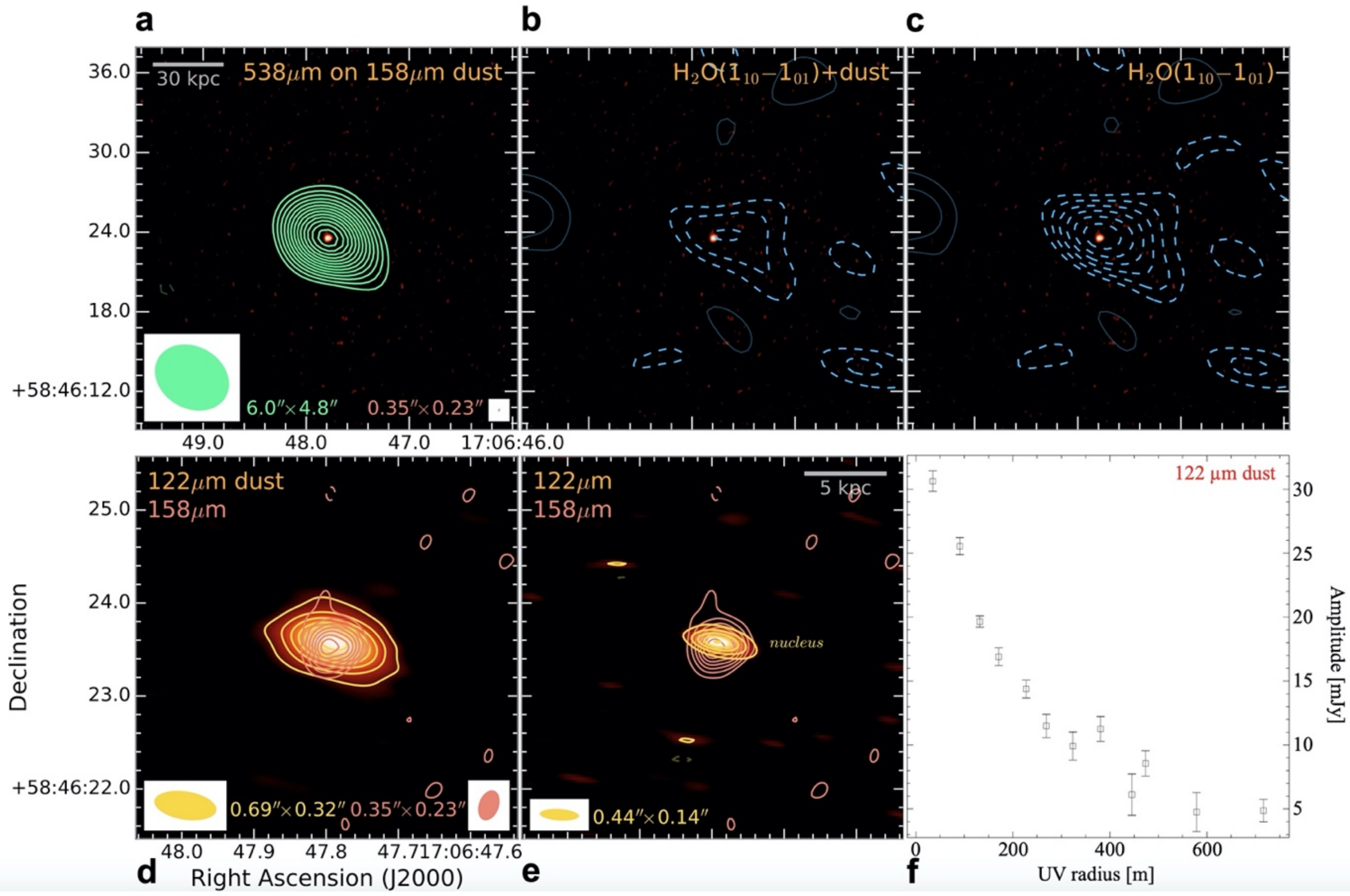

**Extended Data Figure 2: $H_2O$ line emission integrated moment 0 and continuum maps of HFLS3.** **a-c,** $H_2O$ contour maps (blue) before (**b**) and after (**c**) continuum subtraction, and local continuum (**a**, green contours) at the wavelength of the $H_2O$ line, overlaid on 158 µm continuum (intensity scale).[5] $H_2O$ emission is integrated over the central 395 km $s^{-1}$ (100 MHz). **d-e,** Rest-frame 122 µm continuum emission (orange contours and intensity scale) as a proxy for the 108 µm continuum size, showing the full emission (**d**), and the compact nuclear region that accounts for 2/3 of the emission at higher resolution (**e**), overlaid with 158 µm contours (red) for orientation. **f,** Radially-averaged visibility amplitude as a function of interferometer baseline length for the data in **(d)** and **(e)**. The radial profile of the visibility amplitude (binned to 50 m steps, with 1σ error bars) shows that the 122 µm dust emission is clearly resolved. Observed-frame 538 µm continuum contours (**a**) are shown in steps of 1σ=22.5 µJy $beam^{-1}$, starting at +/-3σ. $H_2O$ contours (**b, c**) are shown in steps of 1σ=0.0375 Jy $kms^{-1}$ $beam^{-1}$, starting at +/-2σ. 122 µm contours (**d/e**) are shown in steps of +/10σ/5σ, where 1σ=229/374 µJy $beam^{-1}$, respectively. 158 µm contours (**d, e**) are shown in steps of 3σ, starting at +/-5σ, where 1σ=400 µJy $beam^{-1}$ (all uncertainties are r.m.s.). Negative intensity contours are dashed.

## Extended Data Table 1: Summary of $T_{CMB}$(z) Measurements

| Name | Redshift | Method* | $T_{CMB}(z)$ [K]† | $T_{CMB}(z)/(1+z)$ [K] |
|---|---|---|---|---|
| COBE/FIRAS[7] | 0. | CMB | 2.72548 ± 0.00057 | |
| Planck clusters[17] | 0.037 | $N$=43, tSZ | 2.888±0.039±0.011 | 2.78±0.04±0.01 |
| (dz=0.05 stacks) | 0.072 | $N$=125 | 2.931±0.017±0.011 | 2.73±0.02±0.01 |
| | 0.125 | $N$=92 | 3.059±0.032±0.012 | 2.72±0.03±0.01 |
| | 0.171 | $N$=104 | 3.197±0.030±0.012 | 2.73±0.03±0.01 |
| | 0.220 | $N$=95 | 3.288±0.032±0.013 | 2.70±0.03±0.01 |
| | 0.273 | $N$=87 | 3.416±0.038±0.013 | 2.68±0.03±0.01 |
| | 0.322 | $N$=81 | 3.562±0.050±0.014 | 2.69±0.04±0.01 |
| | 0.377 | $N$=50 | 3.717±0.063±0.014 | 2.70±0.05±0.01 |
| | 0.428 | $N$=45 | 3.971±0.071±0.015 | 2.78±0.05±0.01 |
| | 0.471 | $N$=26 | 3.943±0.112±0.015 | 2.68±0.08±0.01 |
| | 0.525 | $N$=20 | 4.380±0.119±0.016 | 2.87±0.08±0.01 |
| | 0.565 | $N$=18 | 4.075±0.156±0.016 | 2.60±0.10±0.01 |
| | 0.619 | $N$=12 | 4.404±0.194±0.016 | 2.72±0.12±0.01 |
| | 0.676 | $N$=6 | 4.779±0.278±0.017 | 2.85±0.17±0.01 |
| | 0.718 | $N$=7 | 4.933±0.371±0.017 | 2.87±0.22±0.01 |
| | 0.783 | $N$=2 | 4.515±0.621±0.018 | 2.53±0.35±0.01 |
| | 0.870 | $N$=1 | 5.356±0.617±0.019 | 2.86±0.33±0.01 |
| | 0.972 | $N$=1 | 5.813±1.025±0.020 | 2.95±0.52±0.01 |
| Planck clusters[25] | 0.042 | $N$=32, tSZ | 2.856±0.018 | 2.741±0.017 |
| (dz=0.05 stacks) | 0.077 | $N$=186 | 2.953±0.021 | 2.742±0.019 |
| | 0.123 | $N$=114 | 3.072±0.026 | 2.736±0.023 |
| | 0.169 | $N$=83 | 3.162±0.020 | 2.705±0.017 |
| | 0.222 | $N$=46 | 3.326±0.015 | 2.722±0.012 |
| | 0.274 | $N$=20 | 3.495±0.016 | 2.743±0.013 |
| SPT clusters[24] | 0.129 | tSZ | $3.01^{+0.14}_{-0.11}$ | $2.67^{+0.12}_{-0.10}$ |
| | 0.265 | | $3.44^{+0.16}_{-0.13}$ | $2.72^{+0.13}_{-0.10}$ |
| | 0.371 | | $3.53^{+0.18}_{-0.14}$ | $2.57^{+0.13}_{-0.10}$ |
| | 0.416 | | $3.82^{+0.19}_{-0.15}$ | $2.70^{+0.13}_{-0.11}$ |
| | 0.447 | | $4.09^{+0.25}_{-0.19}$ | $2.83^{+0.17}_{-0.13}$ |
| | 0.499 | | $4.16^{+0.27}_{-0.20}$ | $2.78^{+0.18}_{-0.13}$ |
| | 0.590 | | $4.62^{+0.36}_{-0.26}$ | $2.91^{+0.23}_{-0.16}$ |
| | 0.628 | | $4.45^{+0.31}_{-0.23}$ | $2.73^{+0.19}_{-0.14}$ |
| | 0.681 | | $4.72^{+0.39}_{-0.27}$ | $2.81^{+0.23}_{-0.16}$ |
| | 0.742 | | $5.01^{+0.49}_{-0.33}$ | $2.88^{+0.28}_{-0.19}$ |
| | 0.887 | | $4.97^{+0.24}_{-0.19}$ | $2.63^{+0.13}_{-0.10}$ |
| | 1.022 | | $5.37^{+0.22}_{-0.18}$ | $2.66^{+0.11}_{-0.09}$ |
| J085726+185524[23] | 1.7293 | $T_{ex}$(CO) | $7.5^{+1.6}_{-1.2}$ | $2.75^{+0.59}_{-0.44}$ |
| “[27] | | $T_{ex}(CO)^{corr}$ | $6.5^{+1.6}_{-1.2}$ | $2.38^{+0.59}_{-0.44}$ |
| “[3] | | $T_{ex}(CO)^{corr}$ | $7.9^{+1.7}_{-1.4}$ | $2.89^{+0.62}_{-0.51}$ |
| J104705+205734[23] | 1.7738 | $T_{ex}$(CO) | $7.8^{+0.7}_{-0.6}$ | $2.81^{+0.25}_{-0.22}$ |
| “[27] | | $T_{ex}(CO)^{corr}$ | $6.8^{+0.8}_{-0.7}$ | $2.45^{+0.29}_{-0.25}$ |
| “[3] | | $T_{ex}(CO)^{corr}$ | $6.6^{+1.2}_{-1.1}$ | $2.38^{+0.43}_{-0.40}$ |
| Q1331+170[19, 36] | 1.7765 | $T_{ex}$(CI) | $7.2^{+0.8}_{-0.8}$ | $2.59^{+0.29}_{-0.29}$ |
| Q0013-004[20] | 1.9731 | $T_{ex}$(CI) | $7.9^{+1.0}_{-1.0}$ | $2.66^{+0.34}_{-0.34}$ |
| J170542+354340[23] | 2.0377 | $T_{ex}$(CO) | $8.6^{+1.1}_{-1.0}$ | $2.83^{+0.36}_{-0.33}$ |
| “[27] | | $T_{ex}(CO)^{corr}$ | $7.6^{+1.2}_{-1.1}$ | $2.50^{+0.40}_{-0.36}$ |
| “[3] | | $T_{ex}(CO)^{corr}$ | $8.6^{+1.9}_{-1.4}$ | $2.83^{+0.63}_{-0.46}$ |
| B1444+0126[37,3] | 2.0870 | $T_{ex}(CI)^{corr}$ | <10.5 | <3.4 |
| PKS1232+0815[21] | 2.3371 | $T_{ex}(CI)^{corr}$ | $10^{+4}_{-4}$ | $3.0^{+1.2}_{-1.2}$ |
| “[38,3] | | $T_{ex}(CI)^{corr}$ | <9.4 | <2.8 |
| J143912+111740[39] | 2.4184 | $T_{ex}$(CO) | $9.2^{+0.7}_{-0.7}$ | $2.69^{+0.20}_{-0.20}$ |
| “[27] | | $T_{ex}(CO)^{corr}$ | $7.9^{+0.8}_{-0.8}$ | $2.31^{+0.23}_{-0.23}$ |
| “[3] | | $T_{ex}(CO)^{corr}$ | $9.0^{+0.8}_{-0.7}$ | $2.63^{+0.23}_{-0.20}$ |
| “[3] | | $T_{ex}(CI)^{corr}$ | <13.7 | <4.0 |
| J1513+0352[40,3] | 2.4636 | $T_{ex}(CI)^{corr}$ | <12 | <3.5 |
| J000015+004833[28] | 2.5255 | $T_{ex}$(CO) | $9.6^{+0.7}_{-0.6}$ | $2.72^{+0.20}_{-0.17}$ |
| “[27] | | $T_{ex}(CO)^{corr}$ | $7.3^{+0.8}_{-0.7}$ | $2.07^{+0.23}_{-0.20}$ |
| “[41,3] | | $T_{ex}(CO)^{corr}$ | $9.8^{+0.7}_{-0.6}$ | $2.78^{+0.20}_{-0.17}$ |
| “[3] | | $T_{ex}(CI)^{corr}$ | $11.1^{+1.5}_{-6.6}$ | $3.15^{+0.43}_{-1.9}$ |
| J0812+3208[42] | 2.6263 | $T_{ex}(CI)^{corr}$ | $10.8^{+1.4}_{-3.3}$ | $2.98^{+0.39}_{-0.91}$ |
| J123714+064759[43] | 2.6896 | $T_{ex}$(CO) | $10.5^{+0.8}_{-0.6}$ | $2.85^{+0.22}_{-0.16}$ |
| “[27] | | $T_{ex}(CO)^{corr}$ | $9.2^{+0.9}_{-0.7}$ | $2.49^{+0.24}_{-0.19}$ |
| “[3] | | $T_{ex}(CO)^{corr}$ | $10.4^{+0.8}_{-0.7}$ | $2.82^{+0.22}_{-0.19}$ |
| “[3] | | $T_{ex}(CI)^{corr}$ | <13.8 | <3.7 |
| J0843+0221[44,3] | 2.7866 | $T_{ex}(CI)^{corr}$ | <16 | <4.2 |
| Q0347-3819[22] | 3.025 | $T_{ex}$(CII) | $12.1^{+1.7}_{-3.2}$ | $3.01^{+0.42}_{-0.80}$ |
| J2100-0641[45,3] | 3.0915 | $T_{ex}(CI)^{corr}$ | $12.9^{+3.3}_{-4.5}$ | $3.15^{+0.81}_{-1.1}$ |
| J0816+1446[46,3] | 3.2874 | $T_{ex}(CI)^{corr}$ | $15.2^{+1.0}_{-4.2}$ | $3.55^{+0.23}_{-1.0}$ |
| HFLS3 | 6.3369 | $H_2O$ | 16.4-30.2 | 2.24-4.12 |

* CMB: measured from the Cosmic Microwave Background; tSZ: measured from the thermal Sunyaev-Zel'dovich effect in galaxy clusters; $N$: number of clusters contributing to the stacked measurements; $T_{ex}$: indirect measurement based on the excitation temperature of CO, [CI], or [CII] of UV absorption lines along the lines-of-sight to background quasars, with model-based excitation corrections applied where indicated (different values indicate model corrections from different authors to the same data); $H_2O$: Water-based method introduced here.

† Where provided, the second set of error bars indicates systematic uncertainties. Other error bars correspond to statistical uncertainties.